\documentclass[10pt,twocolumn, amssymb, amsmath, aps, showpacs, footinbib, prb]{revtex4-1}
\usepackage{graphicx}

\newcommand{\eff}{\text{eff}}
\newcommand{\AFM}{\text{AFM}}
\newcommand{\FM}{\text{FM}}

\newcommand{\BEC}{\text{BEC}}
\newcommand{\BKT}{\text{BKT}}

\begin{document}

\title{\emph{Ab initio} modeling of Bose-Einstein condensation in Pb$_2$V$_3$O$_9$}

\author{Alexander A. Tsirlin}
\email{altsirlin@gmail.com}
\author{Helge Rosner}
\email{Helge.Rosner@cpfs.mpg.de}
\affiliation{Max Planck Institute for Chemical Physics of Solids, N\"{o}thnitzer
Str. 40, 01187 Dresden, Germany}


\begin{abstract}
We apply density functional theory band structure calculations and quantum Monte Carlo simulations to investigate the Bose-Einstein condensation in the spin-$\frac12$ quantum magnet Pb$_2$V$_3$O$_9$. In contrast to previous conjectures on the one-dimensional nature of this compound, we present a quasi-two-dimensional model of spin dimers with ferromagnetic and antiferromagnetic interdimer couplings. Our model is well justified microscopically, and provides a consistent description of the experimental data on the magnetic susceptibility, high-field magnetization, and field vs. temperature phase diagram. The Bose-Einstein condensation in the quasi-two-dimensional spin system of Pb$_2$V$_3$O$_9$ is largely governed by intralayer interactions, whereas weak interlayer couplings have a moderate effect on the ordering temperature. The proposed computational approach is an efficient tool to analyze and predict high-field properties of quantum magnets.
\end{abstract}

\maketitle

Bose-Einstein condensation (BEC) is one of the fundamental phenomena in physics. Recent activity in the field of quantum magnetism opened new prospects for extensive experimental investigation of this phenomenon.\cite{giamarchi2008} The bosonic nature of magnons and their control by the external magnetic field give rise to simple realization of the BEC, whereas a variety of spin lattices available in transition-metal compounds lead to different regimes of bosonic interactions. However, the detailed understanding of the underlying magnetic couplings remains an essential prerequisite for the correct interpretation of the observed high-field physics. For example, the proposed two-dimensional (2D) regime of BEC in BaCuSi$_2$O$_6$ (Ref.~\onlinecite{sebastian2006}) was based on an oversimplified spin model and later challenged by direct observations of inequivalent spin dimers in this compound.\cite{ruegg2007,kraemer2007} A subsequent theoretical study reconciled all the experimental observations within an extended three-dimensional (3D) model comprising two inequivalent sublattices with different boson densities.\cite{laflorencie2009} 

To derive a complete spin model, inelastic neutron scattering studies are usually required. A viable alternative is given by density functional theory (DFT) band structure calculations that are able to evaluate individual exchange couplings in specific compounds. Here, we show that DFT calculations combined with efficient numerical techniques lead to a quantitative description of the BEC in Pb$_2$V$_3$O$_9$. Based on this approach, the role of individual exchange couplings can be explored, and the regime of the BEC in a quantum magnet can be predicted.

\begin{figure}
\includegraphics{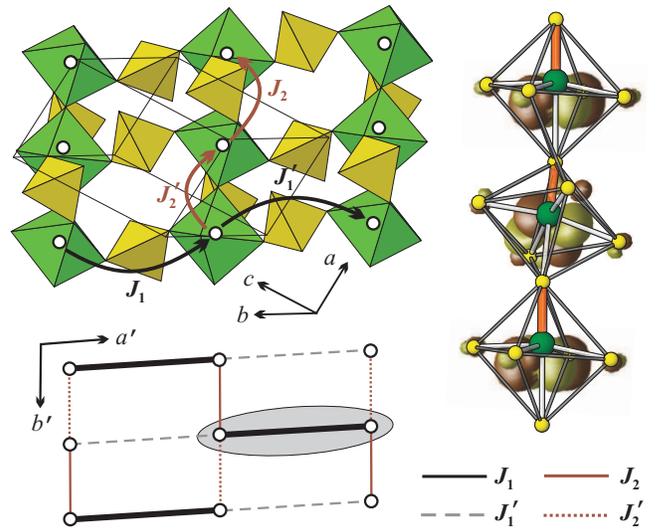}
\caption{\label{structure}
(Color online) Crystal structure (top) and spin model (bottom) of Pb$_2$V$_3$O$_9$. Circles denote positions of V$^{+4}$ and sites of the spin lattice. The shading shows a single spin dimer. The right panel depicts a single chain of corner-sharing VO$_6$ octahedra and respective Wannier functions with $d_{xy}$ orbital character. Thick orange lines show the short V--O bonds.
}
\end{figure}
Pb$_2$V$_3$O$_9$ is a model spin-$\frac12$ material showing magnetic-field-induced long-range ordering (LRO) interpreted as the BEC of magnons.\cite{waki2004,conner2010} Despite a number of experimental studies available, the microscopic magnetic model of this compound remains controversial. The crystal structure comprises chains of corner-sharing V$^{+4}$O$_6$ octahedra (``structural chains'') which are joined into layers by non-magnetic V$^{+5}$O$_4$ tetrahedra (Fig.~\ref{structure}). The magnetic behavior is roughly captured by the model of alternating spin-$\frac12$ chains. Refs.~\onlinecite{waki2004,conner2010} suggest that these spin chains coincide with the structural chains. By contrast, Mentr\'e \textit{et al.}\cite{mentre2008} found alternating spin chains perpendicular to the structural chains. The latter study also proposed ferromagnetic (FM) couplings between the spin chains, although no detailed comparison to the experimental data was given. 

A microscopic investigation of the uniform-spin-chain compound Sr$_2$V$_3$O$_9$ with a similar crystal structure\cite{kaul2003} suggests that leading antiferromagnetic (AFM) couplings run perpendicular to the structural chains. Taking into account the lower symmetry (monoclinic in Sr$_2$V$_3$O$_9$ and triclinic in Pb$_2$V$_3$O$_9$), one would expect AFM alternating spin chains perpendicular to the structural chains and thus support the model by Mentr\'e \textit{et al.}\cite{mentre2008} However, the couplings between the spin chains are also important. Below, we show that Pb$_2$V$_3$O$_9$ should be considered as a quasi-2D system in contrast to the quasi-one-dimensional (1D) spin lattice in Sr$_2$V$_3$O$_9$.

To elucidate the microscopic magnetic model of Pb$_2$V$_3$O$_9$, we performed scalar-relativistic DFT calculations using the \texttt{FPLO} code\cite{fplo} and the local density approximation (LDA) with the exchange-correlation potential by Perdew and Wang.\cite{pw92} The V$^{+4}$-related states were further introduced into a multi-orbital Hubbard model that allowed to treat strong correlation effects in the V $3d$ shell and to evaluate the exchange couplings. The results were cross-checked by a local spin-density approximation (LSDA)+$U$ method with an around-mean-field double-counting correction (DCC) that accounts for electronic correlations in a mean-field fashion. LSDA+$U$ total energies for collinear spin configurations were mapped onto the classical Heisenberg model to obtain individual exchange couplings $J_i$. The $k$-mesh comprised 512 points for the LDA calculation and $50-100$ points for the LSDA+$U$ supercell calculations.

\begin{figure}
\includegraphics{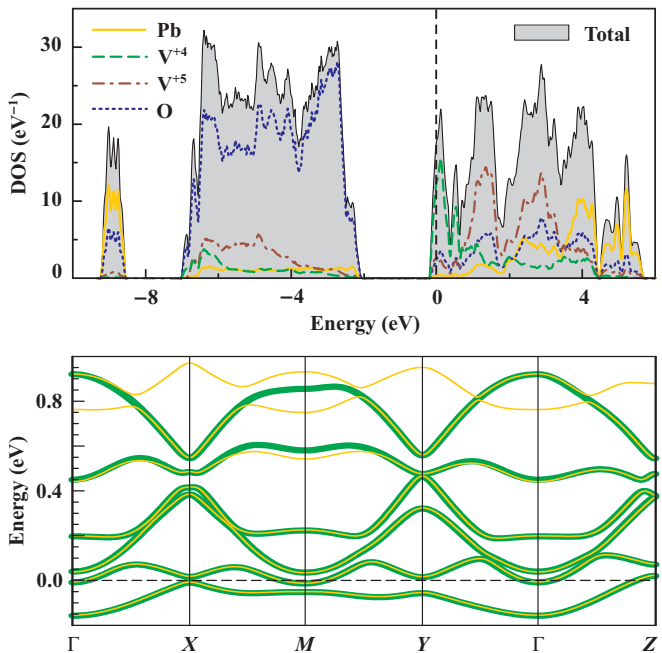}
\caption{\label{dos}
(Color online) Top: LDA density of states for Pb$_2$V$_3$O$_9$. The Fermi level is at zero energy. Bottom: LDA bands at the Fermi level (thin light lines) and the fit with the three-orbital tight-binding model (thick dark lines). The notation of $k$-points is $\Gamma(0,0,0)$, $X(0.5,0,0)$, $M(0.5,0.5,0)$, $Y(0,0.5,0)$, and $Z(0,0,0.5)$ in units of the respective reciprocal lattice parameters.
}
\end{figure}

The LDA band structure of Pb$_2$V$_3$O$_9$ (Fig.~\ref{dos}) resembles that of A$_2$V$_3$O$_9$ with A = Sr and Ba.\cite{kaul2003} The filled valence bands between $-7$~eV and $-2$~eV are formed by O $2p$ states, whereas the bands at the Fermi level originate from $3d$ states of octahedrally coordinated vanadium (i.e., V$^{+4}$). Owing to the complete charge ordering,\cite{mentre2008,mentre1999} the $3d$ states of tetrahedrally coordinated vanadium atoms (i.e., V$^{+5}$) appear above 1~eV only. Lead orbitals give rise to narrow bands around $-9$~eV ($6s$) and to a pronounced contribution above 2~eV ($6p$), similar to other Pb$^{+2}$ compounds.\cite{tsirlin2010} 

V$^{+4}$ bands show a characteristic crystal-field splitting with $t_{2g}$ states lying below 1~eV and $e_g$ states spanning from 1~eV to 4~eV. To extract the relevant microscopic information, we fit the $t_{2g}$ bands with a three-orbital tight-binding model,\cite{note1} based on Wannier functions with proper orbital characters.\cite{wannier} The fit reveals a lower energy of the $d_{xy}$ orbital (0.04~eV) and higher energies of the $d_{yz}$ and $d_{xz}$ orbitals (0.37 and 0.50~eV, respectively) where $z$ denotes the direction of the shortest V--O bond that aligns with the structural chains (Fig.~\ref{structure}). Strong electronic correlations typical of V$^{+4}$ compounds stabilize the half-filled $d_{xy}$ orbital in the Mott-insulating state (e.g., via the LSDA+$U$ calculation). Therefore, the exchange couplings can be estimated using the Kugel-Khomskii model:\cite{kugel1982,mazurenko2006}
\begin{equation}
  J=\dfrac{4t_{xy}^2}{U_{\eff}}- \sum_{\alpha=yz,xz}\dfrac{4t_{xy\rightarrow\alpha}^2J_{\eff}}{(U_{\eff}+\Delta_{\alpha})(U_{\eff}+\Delta_{\alpha}-J_{\eff})},
\label{exchange}\end{equation}
where $t_{xy}$ and $t_{xy\rightarrow\alpha}$ are the hoppings between the $xy$ states and from the $xy$ (half-filled) to $\alpha$ (empty) states, $U_{\eff}$ and $J_{\eff}$ are the effective on-site Coulomb repulsion and Hund's coupling in V $3d$ bands, respectively, and $\Delta_{\alpha}$ is the crystal-field splitting between the $xy$ and $\alpha$ states. The first term in Eq.~\eqref{exchange} corresponds to AFM couplings due to the hoppings between the half-filled $xy$ states. The second term is the FM coupling caused by the hoppings to the empty states and by the Hund's coupling.

Using the typical values of $U_{\eff}=4$~eV and $J_{\eff}=1$~eV,\cite{mazurenko2006,tsirlin2010} we find two AFM and two FM couplings in Pb$_2$V$_3$O$_9$ (Table~\ref{tab_exchange}). The stronger AFM couplings $J_1$ and $J_1'$ run between the structural chains via VO$_4$ tetrahedra, whereas the weaker FM couplings $J_2$ and $J_2'$ run along the structural chains (Fig.~\ref{structure}). This counter-intuitive scenario is readily explained by the orbital state of vanadium. In terms of the crystal-field theory, the energy preference of the $xy$ orbital is caused by the short V--O bond which is directed along the structural chain. The half-filled ($xy$) orbital is therefore located in the plane perpendicular to the structural chain (Fig.~\ref{structure}), and favors the long-range couplings $J_1$ and $J_1'$. On the other hand, it precludes the V--O--V superexchange, despite the V--O--V angles are about $145^{\circ}$, i.e., far from $90^{\circ}$ (see also Refs.~\onlinecite{kaul2003,nath2008}). The resulting couplings along the structural chains ($J_2,J_2'$) are weak and FM due to the short V--V distances. A similar scenario has been established for (VO)$_2$P$_2$O$_7$, Sr$_2$V$_3$O$_9$, and other V$^{+4}$ compounds.\cite{kaul2003,garrett1997,nath2008}

Our model basically follows the earlier proposal by Mentr\'e \textit{et al.}\cite{mentre2008} and differs from Refs.~\onlinecite{waki2004,conner2010}, which assume AFM couplings along the structural chains. In contrast to the result of Mentr\'e \textit{et al.} ($|J_2|$ and $|J_2'|$ are comparable to or even larger than $J_1$ and $J_1'$),\cite{mentre2008} we find rather weak FM couplings that are in agreement with the experiment (see below) and do not inhibit the formation of the spin gap in Pb$_2$V$_3$O$_9$.

\begin{table}
\caption{\label{tab_exchange}
Exchange integrals (in~K) calculated using the model approach [Eq.~\eqref{exchange}] and LSDA+$U$ supercell approach, see text for details. }
\begin{ruledtabular}
\begin{tabular}{ccrrrc}
        & Distance & $J^{\AFM}$ & $J^{\FM}$ &  $J$      &    $J$ \\
        & (\r A)   & \multicolumn{3}{c}{Model approach} & LSDA+$U$ \\\hline
 $J_1$  & 6.28     &  55        &  $-4$     &  51       &  37    \\
 $J_1'$ & 6.32     &  40        &  $-1$     &  39       &  26    \\
 $J_2$  & 3.69     &   0        &  $-2$     &  $-2$     &  $-4$  \\
 $J_2'$ & 3.70     &   0        &  $-1$     &  $-1$     &  $-5$  \\
\end{tabular}
\end{ruledtabular}
\end{table}
The model analysis using Eq.~\eqref{exchange} is confirmed by LSDA+$U$ calculations. Employing the on-site Coulomb repulsion and exchange\cite{note5} parameters $U_d=3$~eV and $J_d=1$~eV, respectively,\cite{tsirlin2009} we find good agreement with the results of the model analysis (Table~\ref{tab_exchange}). The estimated band gap of 1.55~eV conforms to the reported black color of Pb$_2$V$_3$O$_9$.\cite{mentre1999} We also used the fully-localized-limit DCC and the generalized gradient approximation for the exchange-correlation potential, but these changes in the computational procedure did not modify the results qualitatively.

The resulting spin model of Pb$_2$V$_3$O$_9$ is shown in the bottom part of Fig.~\ref{structure}. Although it can be viewed as AFM alternating spin chains (along $a'$) that are coupled ferromagnetically (along $b'$), the numerical study and the comparison to the experiment emphasize the quasi-2D nature of the material (see below). Regarding the interlayer couplings, we note that each V$^{+4}$ site has eight neighbors in the adjacent layers (V--V distances of $8.7-9.2$~\r A), yet most of these contacts are inactive ($|J_i|<0.15$~K). The leading interlayer coupling $J_{\perp}\simeq 0.6$~K provides two bonds at each lattice site and corresponds to the V--V distance of 9.01~\r A. Thus, the interlayer couplings are uniform and unfrustrated. In the following, we apply this model to interpret the magnetic behavior of Pb$_2$V$_3$O$_9$. 

Thermodynamic properties and magnetic ordering temperatures were simulated using directed loop algorithm in the stochastic series expansion representation,\cite{dirloop} as implemented in the ALPS package.\cite{alps} We used 2D and 3D finite lattices with periodic boundary conditions. The lattice size was set to $L\times L$ (2D model) and $L\times L\times L/2$ or $L\times L\times L/4$ (3D model) with $L\leq 32$ (here, $L$ is the number of unit cells; each cell comprises four magnetic atoms).\cite{note2}

\begin{figure}
\includegraphics{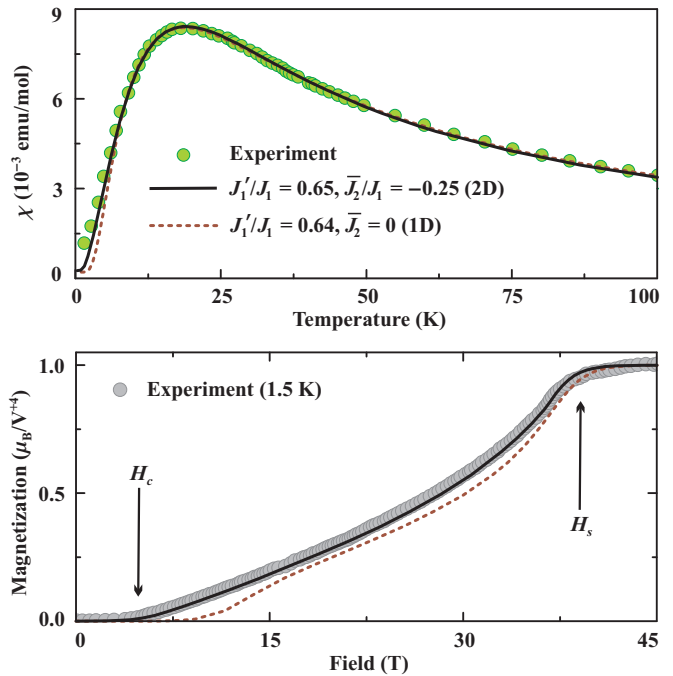}
\caption{\label{chi}
(Color online) Magnetic susceptibility (top) and high-field magnetization (bottom) of Pb$_2$V$_3$O$_9$ (circles) fitted with the 1D alternating-chain model (dashed line) and the 2D \mbox{$J_1-J_1'-\bar J_2$} model (solid line, see also Fig.~\ref{structure}). The magnetization curve was measured at 1.5~K, whereas the model curves are simulated at $T/J_1=0.05$. Experimental data are taken from Ref.~\onlinecite{waki2004}.
}
\end{figure}

Figure~\ref{chi} shows magnetic susceptibility ($\chi$) and high-field magnetization ($M$) of Pb$_2$V$_3$O$_9$ along with the simulations for the purely one-dimensional (1D) alternating spin chain $J_1-J_1'$ model (dashed line: $J_1\simeq 31$~K, $J_1'/J_1\simeq 0.64$, $g\simeq 1.99$). This model underestimates $\chi$ at low temperatures and overestimates the spin gap. Moreover, the shape of the simulated magnetization curve is apparently different from the experimental result while the fitted $g$ value exceeds the actual $g\simeq 1.93$ from electron spin resonance (ESR).\cite{esr} Thus, the alternating-chain model is insufficient to understand the magnetic behavior of Pb$_2$V$_3$O$_9$. 

To improve the 1D model, we consider interchain couplings $J_2$ and $J_2'$. For the moment, we assume a single interchain coupling $\bar J_2=J_2=J_2'$. The effect of inequivalent $J_2$ and $J_2'$ will be discussed later. Since the 1D model yields an accurate prediction of the saturation field, additional (i.e., interchain) couplings should be FM. Indeed, we find a perfect fit of the experimental data with $J_1\simeq 31$~K, $J_1'/J_1=0.65$, $\bar J_2/J_1=-0.25$, and $g\simeq 1.94$ (compare to 1.93 from ESR).\cite{note3} 

The fitted exchange parameters are in remarkable agreement with our DFT results. The leading AFM coupling is $J_1=31$~K (37~K in LSDA+$U$, see Table~\ref{tab_exchange}), whereas the interchain couplings are indeed FM. The interchain coupling $\bar J_2/J_1=-0.25$ is sufficient to reduce the spin gap of Pb$_2$V$_3$O$_9$ and smoothen the magnetization curve with respect to the simulation result for the 1D model. A stronger interchain coupling (as proposed by Mentr\'e \textit{et al.}\cite{mentre2008}) will close the spin gap, thus contradicting the experimental magnetic behavior of Pb$_2$V$_3$O$_9$. 

The key feature of Pb$_2$V$_3$O$_9$ is the field-induced LRO (i.e., BEC) between the critical field $H_c$ and the saturation field $H_s$. Since LRO is a 3D phenomenon, we supply the 2D $J_1-J_1'-\bar J_2$ model with a uniform interlayer coupling $J_{\perp}$. The ordering transition is reflected by a change in the spin stiffness $\rho_s$. For a finite lattice with the fixed aspect ratio, the spin stiffness at the transition temperature $T_{\BEC}$ scales as $\rho_s(T_{\BEC})=L^{2-D}$, where $D=3$ is the dimensionality of the system.\cite{sengupta2003,sengupta2009} Therefore, $L\rho_s(T_{\BEC})$ should not depend on $L$, and the transition temperature can be found as a crossing point of $L\rho_s(T)$ curves calculated for different $L$. Since our system is strongly anisotropic, the aspect ratio of the finite lattices should be adjusted to achieve the proper scaling.\cite{sandvik1999} Following Refs.~\onlinecite{sengupta2003,sengupta2009}, we reduce the lattice size along $J_{\perp}$, and use\cite{note4} $L\times L\times L/2$ for $0.01\leq J_{\perp}/J_1\leq 0.05$ as well as $L\times L\times L/4$ for $J_{\perp}/J_1=0.005$.

\begin{figure}
\includegraphics{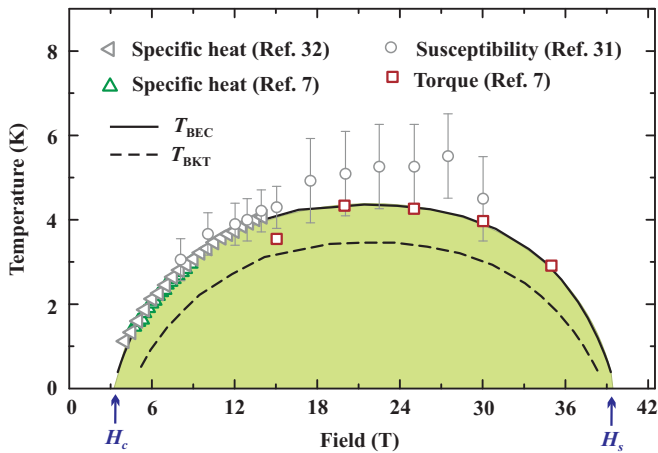}
\caption{\label{diagram}
(Color online) Temperature-vs-field phase diagram for Pb$_2$V$_3$O$_9$. The solid and dashed lines show the temperatures of the BEC (3D model) and BKT (2D model) transitions, respectively. Circles,\cite{waki2007} squares,\cite{conner2010} and triangles\cite{conner2010,nawa2010} represent the experimental data. The shading denotes the region of the BEC phase.
}
\end{figure}
Experimental information on $T_{\BEC}$ of Pb$_2$V$_3$O$_9$ is given by different techniques. The most accurate estimates are available for $\mu_0H\leq 14$~T from specific heat,\cite{waki2004,conner2010,nawa2010} magnetic susceptibility,\cite{waki2004} NMR,\cite{waki2005} and ESR\cite{esr} measurements. In higher fields, $T_{\BEC}$ was probed by magnetic torque\cite{conner2010} and susceptibility.\cite{waki2007} However, the present high-field measurements are subject to significant errors so that the estimates of the maximum $T_{\BEC}$ range from 4.5 to 5.5~K in fields of $20-27$~T (Fig.~\ref{diagram}). In our simulations, we first focused on the region below 14~T and selected $J_{\perp}/J_1=0.005$ (about 0.15~K), which yields the best agreement with the experimental data below 14~T. The value of $J_{\perp}$ supports our DFT result of $J_{\perp}\simeq 0.6$~K. Note that a weak $J_{\perp}$ has nearly no effect on the magnetic susceptibility or magnetization curves, hence the interlayer coupling can be accurately estimated from $T_{\BEC}$ only.

\begin{figure}
\includegraphics{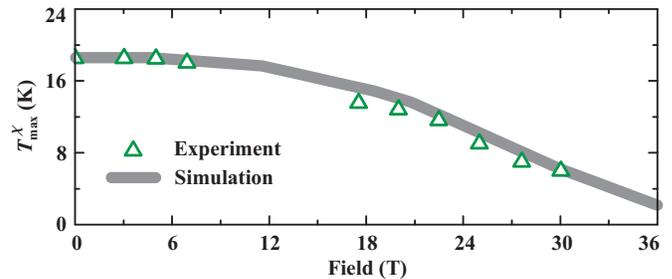}
\caption{\label{maximum}
(Color online) Field-dependent temperature of the magnetic susceptibility maximum ($T_{\max}^{\chi}$) in Pb$_2$V$_3$O$_9$. Triangles represent the experimental data of Ref.~\onlinecite{waki2007}.
}
\end{figure}

The full temperature-vs-field phase diagram is shown in Fig.~\ref{diagram}. Our simulations predict the maximum $T_{\BEC}$ of 4.4~K at $\mu_0H=20$~T in perfect agreement with the results of the torque measurements.\cite{conner2010} In contrast, the high-field susceptibility measurements\cite{waki2007} likely overestimate the transition temperatures. Specific heat or magnetocaloric effect studies above 14~T should further challenge our prediction. We also calculated the temperatures of susceptibility maxima (Fig.~\ref{maximum}) and found excellent agreement with the experimental data of Ref.~\onlinecite{waki2007}.

\begin{figure}
\includegraphics{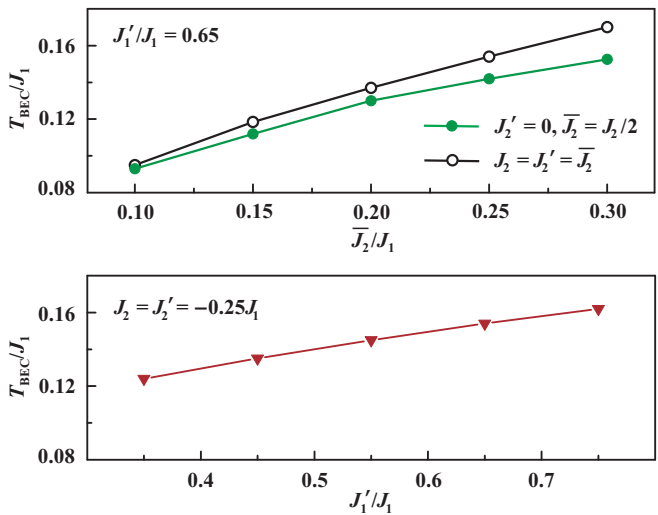}
\caption{\label{trends}
(Color online) The maximum BEC temperature depending on the FM interdimer couplings $J_2$ and $J_2'$ (top) and the AFM interdimer couplings $J_1'$ (bottom). In the upper panel, open and filled circles denote two opposite regimes with $J_2=J_2'$ and $J_2'=0$, respectively. The interlayer coupling is $J_{\perp}/J_1=0.01$.
}
\end{figure}

To understand the origin of the BEC in Pb$_2$V$_3$O$_9$, we first calculate spin-spin correlations in zero field. We find that the system is close to the dimer limit, with the large correlation of $-0.219$ on the $J_1$ bond and weaker correlations of $-0.061$ and $0.031$ on the $J_1'$ and $J_2$ ($J_2'$) bonds, respectively. Magnetic field induces singlet-to-triplet flips on the spin dimers, whereas the interdimer couplings $J_1'$, $J_2$, and $J_2'$ lead to the LRO. The reduction in $J_2$ ($J_2'$) suppresses the BEC (Fig.~\ref{trends}, top). The AFM coupling $J_1'$ has a similar, yet weaker effect (Fig.~\ref{trends}, bottom). The difference should be related to the fact that the FM couplings $J_2$ and $J_2'$ join the dimers into a 2D network, whereas the AFM coupling $J_1'$ connects the dimers into chains and does not lead to the LRO (the $J_1-J_1'$ model is 1D). Based on these results, we identify the FM couplings $J_2$ and $J_2'$ as the main driving force of the BEC in Pb$_2$V$_3$O$_9$. Note also that $J_2$ ($J_2'$) provides an overall energy of $-0.5J_1$ (two bonds per site) which is comparable to $0.65J_1$ arising from $J_1'$. This further justifies the assignment of Pb$_2$V$_3$O$_9$ to quasi-2D systems of coupled spin dimers. Despite leading to a reasonable fit of the magnetic susceptibility, the 1D alternating-chain description is apparently incomplete, and does not capture the essential microscopic physics of the system.

The interlayer coupling $J_{\perp}$ is also important for the BEC. For example, $J_{\perp}/J_1=0.01$ results in the maximum $T_{\BEC}$ of 4.8~K (compare to 4.4~K for $J_{\perp}/J_1=0.005$). The experimental data give a strong evidence for a non-zero $J_{\perp}$, and do not fit to a purely 2D model. In two dimensions, the LRO in a Heisenberg system is possible only at zero temperature, but the magnetic field induces an XY anisotropy giving rise to the finite-temperature Berezinsky-Kosterlitz-Thouless (BKT) transition. Using the scaling procedure from Ref.~\onlinecite{troyer1998}, we estimated $T_{\BKT}$ for the 2D $J_1-J_1'-\bar J_2$ model.\cite{note6} The field dependence of $T_{\BKT}$ (dashed line in Fig.~\ref{diagram}) largely resembles that of $T_{\BEC}$, but the BKT temperatures clearly underestimate the actual transition temperatures in Pb$_2$V$_3$O$_9$. A similar effect has been reported for other quasi-2D systems: spin dimers on a square lattice\cite{laflorencie2009} and a simple square lattice.\cite{sengupta2009} Note that the frustration of interlayer couplings could lead to a peculiar high-field behavior in the vicinity of $H_c$ (Ref.~\onlinecite{laflorencie}). However, our DFT results suggest unfrustrated interlayer couplings in Pb$_2$V$_3$O$_9$.

Finally, we explore the difference between $J_2$ and $J_2'$. To study its effect, we set $J_2/J_1=-0.5$ and $J_2'=0$, thus transferring the full energy of the FM exchange to the $J_2$ bond while keeping the same $\bar J_2/J_1=-0.25$ [here, $\bar J_2=(J_2+J_2')/2$]. No changes in the magnetic susceptibility or magnetization curves are found. By contrast, $T_{\BEC}$ is slightly reduced (compare open and filled circles in Fig.~\ref{trends}). The decrease in $T_{\BEC}$ can be understood as the reduction in the coordination number (two for $J_2=J_2'$ and one for $J_2'=0$) and the subsequent enhancement of quantum fluctuations. Otherwise, the distribution of exchange energies between different bonds of the lattice has only a minor effect on the magnetic behavior as long as the quasi-2D nature of the system is preserved. This situation is typical and can be compared to the effect of spatial anisotropy on a frustrated square lattice\cite{tsirlin2009} or honeycomb lattice.\cite{cu2v2o7}

The proposed magnetic model of Pb$_2$V$_3$O$_9$ is supported by a direct comparison to the experimental data, and discloses the microscopic physics of this material. Although the alternating $J_1-J_1'$ chains form a backbone of the spin lattice, the system should be rather viewed as 2D, since the purely 1D model overestimates the spin gap and does not account for comparable interdimer correlations along $a'$ and $b'$. The 2D model of coupled spin dimers provides a realistic description of Pb$_2$V$_3$O$_9$ by reproducing the temperature dependence of the magnetic susceptibility, the field dependence of the magnetization, and the field dependence of the susceptibility maximum. To achieve an accurate description of the field vs. temperature phase diagram, the weak interlayer coupling $J_{\perp}$ is required.

The quasi-2D nature of Pb$_2$V$_3$O$_9$ should inhibit the observation of effects that are typical for 1D systems. In particular, manifestations of the Luttinger liquid physics, predicted for an isolated alternating spin chain,\cite{sakai1996} are likely concealed by the BEC transition. The distinct features of Pb$_2$V$_3$O$_9$ are the simple spin lattice, the pronounced two-dimensionality, and the combination of FM and AFM interdimer couplings. Most of the known BEC materials reveal a complex arrangement of interdimer couplings (e.g., in TlCuCl$_3$)\cite{cavadini2001,*oosawa2002} or inequivalent spin dimers (e.g., in BaCuSi$_2$O$_6$),\cite{ruegg2007} thus realistic modeling is a tough problem. In contrast, Pb$_2$V$_3$O$_9$ demonstrates the BEC physics on a relatively simple and well-defined quasi-2D spin lattice and on an experimentally accessible field scale. We expect that the recent progress in crystal growth\cite{conner2010,nawa2010} will stimulate extensive investigation of Pb$_2$V$_3$O$_9$ in fields up to $H_s$. The quasi-2D nature of the spin lattice raises the issue of the actual dimensionality, and calls for theoretical study of critical exponents. We also note that our model is presently restricted to isotropic exchange couplings. Since Sr$_2$V$_3$O$_9$ reveals Dzyaloshinsky-Moriya couplings,\cite{kaul2003,ivanshin2003} a sizable exchange anisotropy is also possible in Pb$_2$V$_3$O$_9$ and should be further probed by single-crystal ESR experiments.

In summary, we have studied the electronic structure of Pb$_2$V$_3$O$_9$ and presented the microscopic magnetic model for the BEC in this quantum magnet. The compound can be understood as a quasi-2D system of spin dimers with both antiferromagnetic and ferromagnetic interdimer couplings. Quantum Monte Carlo simulations of the magnetic susceptibility, high-field magnetization, and BEC transition temperatures are in excellent agreement with the experimental data, and yield accurate estimates of individual exchange couplings. The simple and well-characterized spin lattice of Pb$_2$V$_3$O$_9$ makes this compound a convenient model system for future studies.
\medskip

We acknowledge fruitful discussions with Oleg Janson, Franziska Weickert, and Nicolas Laflorencie. A.T. was funded by the Alexander von Humboldt foundation.


\begin{thebibliography}{41}
\bibitem [{not(){\natexlab{c}}}]{note2}%
  \BibitemOpen
  \bibinfo {note} {Note that the unit cell comprises two bonds along $a'$ and
  $b'$, yet there is only one bond in the interlayer direction.}\BibitemShut
  {Stop}%
\bibitem [{not(){\natexlab{d}}}]{note3}%
  \BibitemOpen
  \bibinfo {note} {The remaining discrepancy in $\chi$ below 4~K may be related
  to trace amounts of paramagnetic impurities.}\BibitemShut {Stop}%
\bibitem [{not(){\natexlab{e}}}]{note4}%
  \BibitemOpen
  \bibinfo {note} {Once the proper aspect ratio is found, the error bar for
  $T_{\BEC}$ is below $5$~\% within the temperature range under
  investigation.}\BibitemShut {Stop}%
\bibitem [{not(){\natexlab{f}}}]{note6}%
  \BibitemOpen
  \bibinfo {note} {Specifically, we plot $1/[(\pi\rho_s/T)-2]-\log L$ vs. $L$.
  This quantity converges to a finite value at $T=T_{\BKT}$, and diverges for
  $T>T_{\BKT}$ or $T<T_{\BKT}$.}\BibitemShut {Stop}%
\end{thebibliography}
%

\end{document}